\documentclass[12pt]{iopart}
\usepackage{subcaption}
\usepackage{graphicx}% Include figure files

\begin{document}
\title[Effect of gallium termination on wetting layer properties in InGaAs quantum dots]{Effect of gallium termination on InGaAs wetting layer properties in droplet epitaxy InGaAs quantum dots}

\author{D Fricker$^{1,5}$, P Atkinson$^2$, M Lepsa$^{1,3}$,  Z Zeng$^{1,5}$, A Kovács$^4$, \\
L Kibkalo$^4$, RE Dunin-Borkowski$^{4,5}$, and BE Kardyna\l{}$^{1,5}$}

\address{$^1$ Peter Grünberg Institute 9, Forschungszentrum Jülich, 52425 Jülich, Germany }
\address{$^2$ Institut des Nano Sciences de Paris, CNRS UMR 7588, Sorbonne Université, 75005 Paris, France}
\address{$^3$ Peter Grünberg Institute 10, Forschungszentrum Jülich, 52425 Jülich, Germany }
\address{$^4$ Ernst Ruska-Centre for Microscopy and Spectroscopy with Electrons, Peter Grünberg Institute 5, Forschungszentrum Jülich, 52428 Jülich, Germany }
\address{$^5$ Department of Physics, RWTH Aachen University, 52074 Aachen, Germany}
\ead{\mailto{d.fricker@fz-juelich.de , b.kardynal@fz-juelich.de} }

\begin{abstract}
Self-assembled quantum dots based on III-V semiconductors have excellent
properties for applications in quantum optics. However, the presence of a 2D 
wetting layer which forms during the Stranski-Krastanov growth of 
quantum dots can limit their performance. Here, we 
investigate wetting layer formation during quantum dot growth by 
the alternative droplet epitaxy technique. We use a combination of 
photoluminescence excitation spectroscopy, lifetime measurements, and 
transmission electron microscopy to identify the presence 
of an InGaAs wetting layer in these droplet epitaxy quantum dots, even in the absence of distinguishable wetting layer photoluminescence. We observe that increasing 
the amount of Ga deposited on a GaAs (100) surface prior to the growth 
of InGaAs quantum dots leads to a significant reduction in the 
emission wavelength of the wetting layer to the point where it can no 
longer be distinguished from the GaAs acceptor peak emission in 
photoluminescence measurements.
\end{abstract}
\maketitle
%\ioptwocol

\section{Introduction}\label{sec: Introduction}
The field of semiconductors as a platform for non-classical light sources has been rapidly evolving over the years as a result of growing interest in quantum communication. Photoluminescence from a single InAs/GaAs quantum dot (QD) grown by molecular beam epitaxy (MBE) was first reported in 1994 \cite{Bastard94}. These QDs were grown in the Stranski-Krastanov (SK) growth mode, in which quantum dot formation is driven by strain relaxation of the InAs layer when the layer thickness exceeds a critical thickness \cite{D.Leonard.}. These SK QDs have been a workhorse of semiconductor quantum optics and used extensively for the fabrication of light sources for quantum communication \cite{Intallura07}. Since then, numerous remarkable optical phenomena have been observed in semiconductor-based QDs as their fabrication has been continually improved to achieve better control over QD formation \cite{Lodahl15}. However, one disadvantage of the SK growth mode is the presence of a wetting layer (WL), which leads to decoherence processes in the QDs that are highly undesired in some applications \cite{Seravalli09, Wang.2005, Lobl.2019b}. One method to decrease the influence of this wetting layer on the QDs is to use an AlGaAs capping layer which eliminates confined electronic states in the conduction band in the wetting layer \cite{Lobl.2019b}. \\
In recent years, however, a different growth mode, the droplet epitaxy (DE) technique, has been developed, yielding QDs with comparable optical properties as SK-grown QDs \cite{Gurioli.2019,Tina20}. This technique has also been reported to result in wetting-layer free QDs \cite{Skiba17}. \\
In this report, we show that the emission wavelength of the WL can be controlled by the amount of Ga used to prepare a Ga-terminated surface prior to the growth of InGaAs dots by droplet epitaxy. We study wetting layer electronic states using photoluminescence excitation (PLE) and time-resolved photoluminescence (TR PL). This has allowed the presence of a WL to be identified even when the WL emission cannot be clearly identified in photoluminescence (PL) spectroscopy. The In distribution in the wetting layers grown on different Ga-terminated surfaces is studied using transmission electron microscopy (TEM). 
 \section{ Epitaxy}\label{sec:Epitaxy}
 The investigated samples were grown on undoped (100) GaAs substrates in a molecular beam epitaxy system from MBE Komponenten, using an arsenic valve cracker cell for $As_4$ flux control and a pyrometer to measure the substrate temperature. After thermal deoxidation of the substrate, a  500\,nm thick GaAs buffer layer was grown at 630\,$^\circ$C with a growth rate of 1 monolayer\,(ML)/s. A 120\,nm thick AlAs/GaAs superlattice was incorporated in the middle of the GaAs buffer layer in order to smooth the surface. Following buffer layer growth, the surface was prepared for QD growth. Firstly, the substrate temperature was reduced to 575\,$^\circ$C and then the $As_4$ valve was closed for 5 minutes. 
 Afterwards, the substrate was slowly cooled to 350\,$^\circ$C over 20 minutes to allow the residual arsenic in the chamber to be pumped away. When the chamber background pressure reached 
8$\times$10$^{-9}$\,mbar, 1.8\,ML of Ga was deposited on the GaAs surface with a low flux of 0.06\,ML/s, which should result in a Ga-terminated GaAs surface \cite{Mano.1999, Sanguinetti.2003, Jo.2011}. On this surface 1.4\,ML of In was then deposited with a flux of 0.045\,ML/s to form In droplets. The Ga and In growth rates and layer thicknesses are given as equivalent for GaAs and InAs, respectively. The droplets were immediately crystallized with an $As_4$ beam flux of 2$\times$10$^{-5}$\,mbar to form InGaAs QDs. Simultaneously, slow heating of the wafer was started. Then, after 30 minutes, when the substrate temperature reached 530\,$^\circ$C the QDs were capped with a 2\,nm thick GaAs capping layer, followed by the growth of a 98\,nm thick GaAs layer at 580\,$^\circ$C. Both capping layers were grown with a growth rate of 1\,ML/s. \\
The deposition amounts given here (1.8\,ML Ga, 1.4\,ML In) correspond to the deposition amount at the wafer centre. However, the wafers were grown without substrate rotation during the Ga and In deposition steps, resulting in an expected $\pm 25\,\%$ variation in deposition amount across a 4-inch wafer \cite{MBE}. After Ga deposition, the substrate was rotated to ensure that the In flux gradient was at 90$^\circ$ to the Ga flux gradient. This technique allows the effect of different Ga amounts during the Ga-termination step, without changing other growth parameters, to be studied on a single wafer.\\
To check that the observed variation across the wafer was not due to a temperature gradient, a second wafer with identical growth conditions but a nominal Ga amount of 2.16\,ML was grown. Later, it will be shown that the emission near the centre of the 2.16\,ML Ga wafer (wafer 2) overlaps well with the emission near the edge of the 1.8\,ML Ga wafer (wafer 1) where the Ga amount was $\sim 2.2\,ML$ Ga. This indicates that a temperature gradient from the centre to the edge of the wafer cannot account for the observed shift in emission wavelength.

\section{Sample characterization}\label{sec:Sample characterisation}
\subsection{Optical characterization}\label{subsec:Optical characterization}
Mapping of the PL across the wafer was carried out with meV resolution, using a cold finger He cryostat mounted on a manual stage allowing large movements of the sample.\\
More detailed $\mu$-PL, PLE and TR PL measurements were carried out in a closed-cycle cryostat, where the samples were mounted on a computer-controlled xyz-stage offering sub-$\mu m$ resolution. An achromatic lens with a numerical aperture (NA) of 0.81 focused the laser beam on the sample and collected the resulting PL signal. A continuous wave (CW) Ti:sapphire laser with a tunable wavelength was used to perform the $\mu$-PL and PLE measurements.\\
 To perform the TR PL measurements, a  pulsed diode laser with a tuneable pulse repetition rate up to 80\,MHz and an emission wavelength of 660\,nm was used. The laser pulse width was $\sim 200\,ps$. The emitted PL was spectrally selected with wavelength-tunable filters and guided to a single-photon avalanche photodiode (SPAD), which has 30\,ps time resolution.\\

\subsection{Structural characterization}\label{subsec:Structural characterization}
Structural characterization of the samples was carried out using high-angle annular dark-field (HAADF) scanning TEM (STEM) imaging. Cross-sectional specimens were prepared using focused Ga ion beam (FIB) sputtering in a dual beam scanning electron microscope. The surface damage was reduced by low-energy ($<1$\,keV) Ar ion beam sputtering \cite{Kovacs.2017}. The images were recorded using an electron probe aberration corrected microscope operated at 200\,kV \cite{Kovacs.2016}. The inner annular dark-field detector semi-angle used was 69\,mrad, resulting in HAADF imaging, which is sensitive to chemical composition as the image intensity scales approximately as $I \sim Z^2$, where Z is the atomic number ($Z_{In}=49, Z_{Ga}=31$). 

\begin{figure*}
	\centering
	\begin{subfigure}[t]{5cm}
		\includegraphics[width = 5cm ]{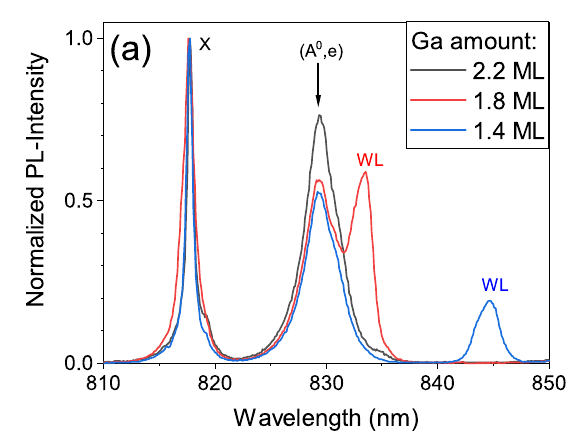}					
	\end{subfigure}
	\begin{subfigure}[t]{5cm}
			\includegraphics[width = 5cm ]{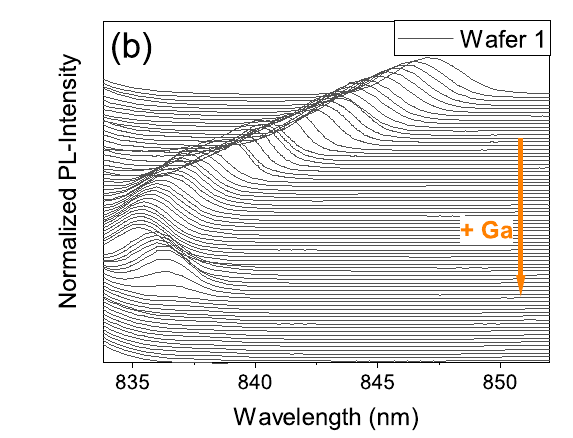}					
	\end{subfigure}
	\begin{subfigure}[t]{5cm}
			\includegraphics[width = 5cm ]{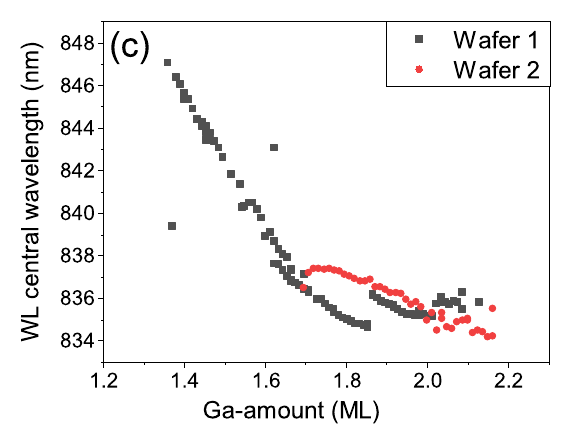}
				
	\end{subfigure}
	\caption{(a) Normalize$ \mu$-PL spectra from three different positions on wafer 1 along the Ga gradient. (b) Normalized wetting layer signal measured on wafer 1 along the Ga gradient in 1\,mm steps. (c) Central wavelength of the Gaussian function fitted to the signal assigned to wetting layer emission for wafers 1 and 2. Wafer 1 had a deposition of 1.8 ML\,Ga at the wafer centre, while wafer 2 had a deposition of 2.16\,ML Ga at the centre. Both wafers had the same deposition amount of In (1.4\,ML) at the centre.}
	\label{fig:1}
\end{figure*}

\section{Results and discussion}\label{sec:Results}
Figure \ref{fig:1}(a) shows normalized $\mu$-PL spectra obtained at different positions on the wafer along the Ga gradient. The positions correspond to Ga amounts of 1.4\,ML, 1.8\,ML and 2.2\,ML. Thanks to the use of the gradient technique during QD layer formation, these three spectra were taken from different positions along the Ga gradient from wafer 1. The deposited InAs amount was 1.4\,ML. The two PL peaks at 818\,nm and 830\,nm originate from radiative recombination of the GaAs free exciton and the free electron-neutral acceptor  ($(A^0,e)$), respectively \cite{Kudo1986, D.Bimberg.}. The positions of these GaAs PL lines are, as expected, independent of the Ga amount deposited during the Ga-termination step.\\ 
A decrease in emission wavelength of the WL peak can be observed when the Ga amount increases from 1.4\,ML to 1.8\,ML. When the Ga amount is 2.2\,ML, the wetting layer peak can no longer be resolved in the PL spectrum. PL measurements in fine steps along the Ga gradient track the development of the WL emission wavelength with Ga amount in more detail.  The spectra shown in Figure \ref{fig:1}(b) were recorded in 1\,mm steps along the Ga-gradient on wafer 1, with the deposited In amount fixed at 1.4\,ML InAs. Each spectrum in Figure \ref{fig:1}(b) was fitted with Gaussian functions. The central wavelengths of the fitted functions are plotted as black points in Figure \ref{fig:1}(c). The position on the wafer is converted into the expected amount of Ga. The WL emission wavelength continuously decreases from 847 to 834\,nm, with the Ga amount increasing from 1.35\,ML to 1.85\,ML. At around 1.87\,ML the WL emission wavelength changes abruptly from 834\,nm to 836\,nm, and remains in the 836$\pm$\,nm range until no WL is observed for larger amounts of Ga of 2.2\,ML. The central wavelength of WL emission on wafer 2 is shown in red in Figure \ref{fig:1}(c) and shows similar behaviour but with no abrupt change in WL emission wavelength at any point. There is a good overlap of the WL emission for wafers 1 and 2 for the  Ga deposition amount between 1.9-2.1\,ML, despite the fact that these Ga deposition amounts occur at different spatial positions on the two wafers. This means that any effect of a difference in substrate temperature with position on the wafer can be neglected, so that the effect of the Ga deposition amount on WL formation is considered in the following discussion only. \\
There are three possible explanations for the absence of a WL PL signal. Firstly, it is possible that the formation of the WL has been suppressed by the Ga deposition amount. It has previously been demonstrated that, on an arsenic-terminated surface,  the first 0.75-1.75\,ML of group III deposition bonds to excess arsenic on the surface, forming a wetting layer and droplet formation only starts to occur once the surface is metal-rich \cite{Sanguinetti.2003}. However, under our growth conditions and with the residual arsenic overpressure in our MBE chamber, it is possible that the transition to a fully Ga-terminated surface was achieved \cite{Jo.2011} only once 2.2\,ML Ga was deposited. It could be expected that all of the deposited In on a Ga-terminated surface forms droplets and no InAs WL is formed.\\
A second possible explanation is that the wetting layer emission persists but overlaps with the acceptor signal, so that differentiation in PL measurements is not possible for samples with more than 2.2\,ML of deposited Ga.  The two PL signals can however be distinguished by time-resolved PL, since WL and acceptor-mediated PL occur with different decay times.\\
The third possible explanation for the absence of the WL PL signal is that there is rapid relaxation of photo-excited carriers from the wetting layer into the QDs, so that PL signal from the wetting layer is not observable. The presence of a wetting layer in the latter case can be probed using PLE. 
\subsection{PLE} \label{subsec:PLE}
\begin{figure*}
	\centering
	\begin{subfigure}[t]{5cm}
		\includegraphics[width=5cm]{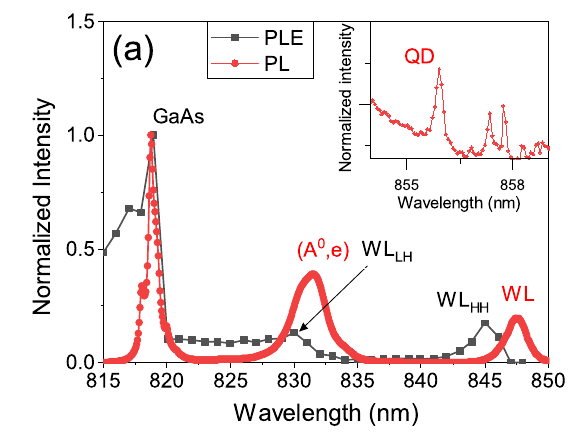}	
	\end{subfigure}
	\begin{subfigure}[t]{5cm}
		\includegraphics[width=5cm]{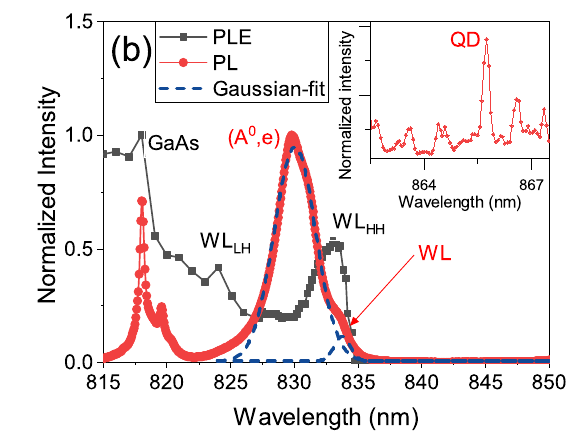}	
	\end{subfigure}
	\begin{subfigure}[t]{5cm}
		\includegraphics[width=5cm]{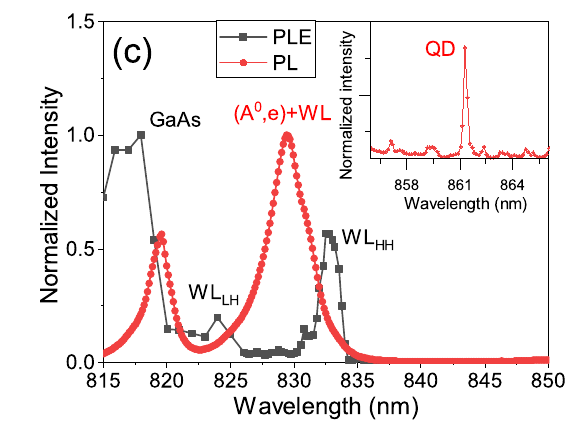}	
	\end{subfigure}
	\caption{Normalized PL and PLE intensities for (a) 1.4\,ML, (b) 2.2\,ML  and (c) 2.5\,ML of deposited Ga amounts for Ga termination. The identified states for the PL data are the GaAs free exciton, $(A^0,e)$ and WL recombinations, whereas the PLE data identifies GaAs, $WL_ {LH}$ and $WL_ {HH}$ states. The inset of each figure shows the PL spectrum of the QD used for the PLE measurements.}
	\label{fig:2}
\end{figure*}

PLE measurements were performed with a wavelength-tunable Ti:Sapphire laser in the $\mu$-PL set-up. The PLE data are shown in Figure \ref{fig:2} in black. The data represent the normalized QD emission intensity as a function of excitation wavelength on three different positions on the wafer. At each position, the PLE spectrum of several QDs, with QD emission wavelength ranging from 855\,nm to 890\,nm, were studied. No significant differences were observed in their PLE spectra, so the data are shown for a single representative QD. The PL of the sensing QD is shown in the inset. PLE spectra are shown together with PL spectra from the same position and spectral range, measured using an excitation wavelength of 780\,nm. \\
Figure \ref{fig:2}(a) shows the data for 1.4\,ML of deposited Ga amount from wafer 1. The PLE data demonstrate a clear increase in QD intensity when the laser light is resonant with the GaAs bandgap at 818\,nm and with the wetting layer states. Among the wetting layer states, both the light hole $WL_{LH}$ and the heavy hole $WL_{HH}$ states are observed \cite{Moskalenko.2002}. The $WL_{HH}$ state is also observable in PL measurements and there is a Stokes shift of 2\,nm (3.5\,meV) between the PLE and PL peak positions. Figures \ref{fig:2}(b) and (c) show the data for 2.2\,ML and 2.5\,ML of Ga from wafer 2, respectively. For 2.2\,ML, the PL from WL emission at a wavelength of 833\,nm partially overlaps with the $(A^0,e)$ signal, which can be seen from fitting with two Gaussian functions. The 2.5\,ML sample does not show a detectable wetting layer in the PL spectrum. The PLE data nevertheless show a clear WL peak at 833\,nm. Thus, it can be shown that a WL exists even though there is no distinguishable WL PL signal. 
 Both the Stokes shift between the PL and PLE signal of the $WL_{HH}$ states and the wavelength difference between heavy and light hole WL states decreases with the wavelength of the PLE $WL_{HH}$  signal, probably as a result of weaker confinement of the different states within the wetting layer. Other PLE measurements on wafers 1 and 2 at positions with different WL wavelengths show the same effect as that presented here. 
\subsection{TR PL}\label{subsec:TR-PL}
\begin{figure}
	\centering
		\includegraphics[width=8cm]{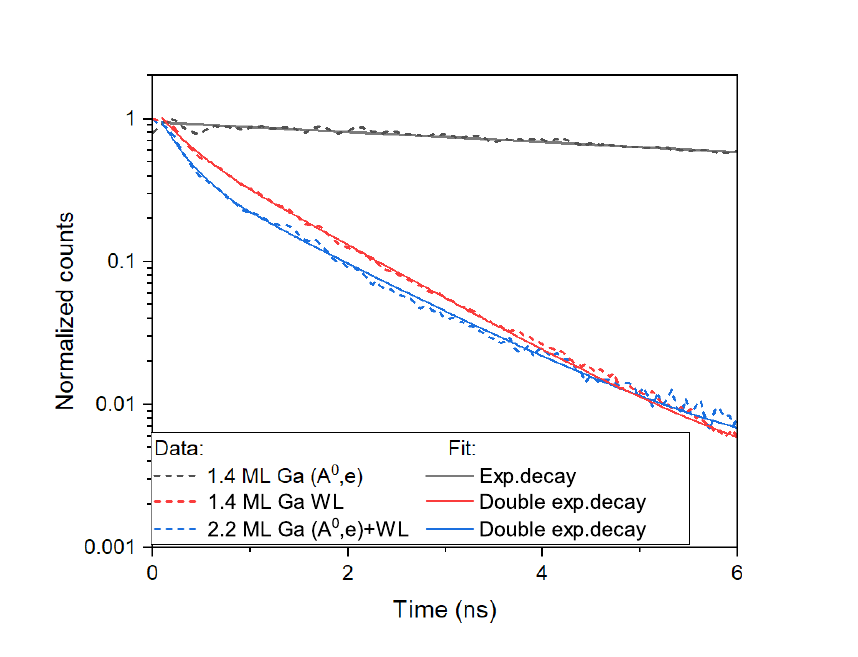}	
	\caption{ Normalized lifetime measurements for a 1.4\,ML Ga sample with spectrally separable  $(A^0,e)$ (black) and WL (red) PL signals with corresponding exponential fits and for the 2.2\,ML Ga with spectrally overlapping $(A^0,e)$ and WL signals (blue), respectively. The signal that contains WL emission is fitted with a double exponential decay and shows significantly shorter lifetimes than pure  $(A^0,e)$ emission.  }
	\label{fig:3}
\end{figure}

Time-resolved photoluminescence measurements were performed in the  $\mu$-PL setup using a pulsed diode laser.  Figure \ref{fig:3} shows lifetime measurements on two samples with different Ga amounts: a 1.4\,ML Ga sample from wafer 1, where the acceptor (in black) and WL (in red) signals are spectrally well separated (as seen in Figure \ref{fig:2}(a)) and a 2.2\,ML Ga sample from wafer 2 (in blue), where the WL cannot be spectrally well separated by PL measurements (shown in Figure \ref{fig:2}(b)). Exponential decay fitting to the 1.4\,ML sample shows that the decay time of $(A^0,e)$ is $\tau_{(A^0,e)} = 13$\,ns, which is significantly greater than the wetting layer lifetimes. The latter is found by fitting double exponential decay functions to be $\tau_{WL_1} = 0.3$\,ns and $\tau_{WL_2} = 1.2$\,ns.  The fast WL decay time may due to exciton relaxation to the QD or other losses, while the slower WL decay time is likely to correspond to the radiative emission from the WL, which is visible in the PL measurement. Similar measurements (not shown here) at different positions on wafers 1 and 2, where the WL can be resolved in the PL spectrum, result in lifetimes in the range $\tau_{(A^0,e)} = 9-13$\,ns, $\tau_{WL_1} = 0.2-0.6$\,ns and $\tau_{WL_1} = 1.2-1.8 $\,ns. It is therefore clear that the wetting layer emission can be distinguished from the acceptor signal due to the significant difference in decay time. \\
The fitting of a double-exponential decay function to the mixed $(A^0,e)$+WL emission of the 2.2\,ML sample gives $\tau_{Mix_1} = 0.3$\,ns and $\tau_{Mix_2} = 1.3$\,ns. Both lifetimes extracted from the decay of the single emission line seen in the 2.2\,ML sample ,$\tau_{Mix}$ (0.3\,ns and 1.3\,ns ), are similar to  the lifetimes of the WL radiative emission seen in the 1.4\,ML sample, $\tau_{WL}$ (0.3\,ns and 1.2\,ns).  In the mixed state the long lifetime of the acceptor excitation $\tau_{(A^0,e)}$ (13\,ns) could not be found.  Further, measurements (not shown here) result in lifetimes of mixed states in the range $\tau_{Mix_1} = 0.2-0.4$\,ns and $\tau_{Mix_2} = 1.2-1.6$\,ns, which is in the range of decay times for resolved WL emission for Ga deposition below 2.2\,ML and is one order of magnitude smaller than that for $(A^0,e)$ emission. These observations indicate that, for Ga deposition larger than 2.2\,ML, where only the acceptor peak appears to be visible in the PL, this peak is in fact composed of emission from $(A^0,e)$ and the WL.

\subsection{TEM}\label{subsec:TEM}
\begin{figure*}	
	\centering
		\includegraphics[width=1\linewidth]{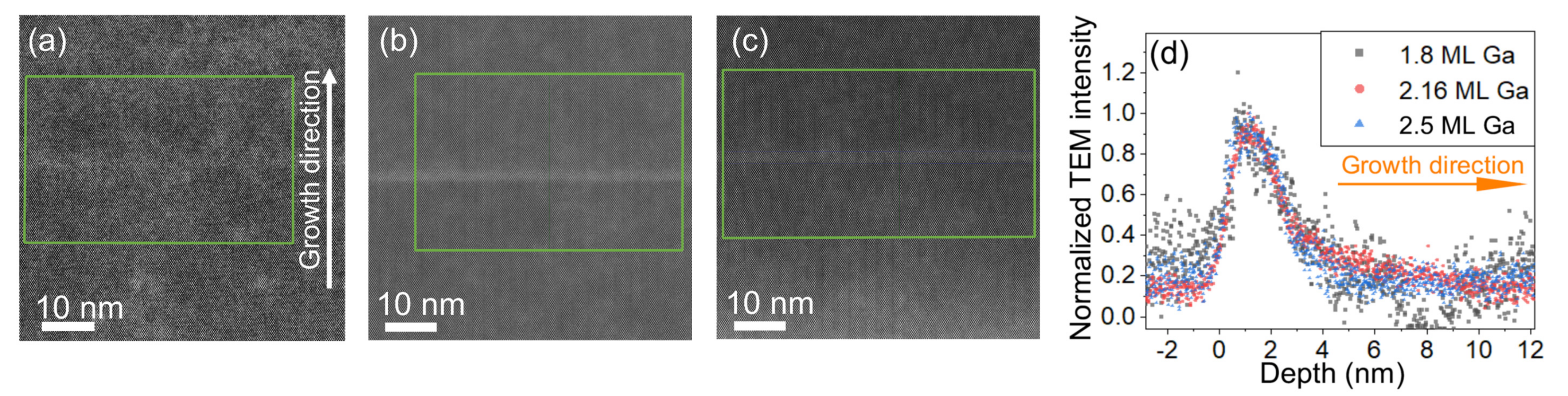}
	\caption{ HAADF STEM images for (a) 1.8\,ML, (b) 2.16\,ML and (c) 2.5\,ML  of deposited Ga amount. The bright horizontal line indicates in all samples a continuous InGaAs layer, corresponding to the wetting layer. d) Normalized line scans for the three samples recorded perpendicular to the WL in HAADF STEM images. }
	\label{fig:4}
\end{figure*}

 Three samples, with deposited Ga amounts of 1.8\,ML Ga, 2.16\,ML Ga and 2.5\,ML Ga, were investigated with TEM. Both samples with higher deposited Ga amounts are from wafer 2. The STEM imaging conditions were identical in studies of the specimens. HAADF STEM images of the three samples are shown in Figure \ref{fig:4}(a-c). In all three samples, a continuous bright line is observed, indicating a continuous InGaAs wetting layer. The sample with the lowest amount of deposited Ga shows the weakest contrast of the wetting layer relative to the surrounding GaAs. This may be due to a small difference in In deposition amount between wafer 1 and 2 despite having nominally the same amount or because of an artefact from the measurements and preparation. Figure \ref{fig:4}(d) shows the normalized line scan intensity (averaged over $\sim$ 70\,nm along the wetting layer) along the growth direction for all three samples.  All distributions show a sharp increase in In signal at the start of the In deposition. The In segregates during overgrowth with GaAs, giving a more gradual drop in In concentration along the growth direction. This profile can be observed for all three samples. The thicknesses of these wetting layers are the same for all Ga deposition amounts. Note that the intensity fluctuation in the linescan of the sample with  1.8\,ML Ga is above the noise in the TEM measurements. \\
 The data presented here provide strong evidence for the presence of a WL in droplet-epitaxy-grown InAs dots on GaAs substrates, even in the case of Ga-terminated surfaces. This may be due to In exchange with Ga atoms on the Ga-terminated surface, resulting in a partially In-terminated surface during the initial stages of In deposition prior to droplet formation. During the recrystallisation step, this In-Ga-terminated surface would form an InGaAs wetting layer. Alternatively, outdiffusion from the In droplet during the recrystallisation step may result in the formation of a thin WL between the dots.  This kind of outdiffusion has been observed with GaAs droplets on AlGaAs surfaces, leading to nano-disks with a diameter of up to several 100 nm forming around the dots \cite{Bietti.2014}. We have not seen evidence for this disk formation when carrying out AFM measurements on surface droplet epitaxy dots grown under the conditions investigated here. However, given the greater diffusion length of In compared to Ga atoms on GaAs \cite{Wakejima.1994, Koshiba.1994}, we cannot exclude that this has occurred but that the ring is so large and small in height that it has not been observed in AFM measurements.

\section{Conclusions}\label{sec:Conclusion}
We have demonstrated that the Ga deposition amount prior to In droplet epitaxy is an effective tool to tune the WL emission wavelength. This emission wavelength was reduced until the WL signal overlapped with the GaAs free-electron-acceptor signal in PL measurements, such that these signals could not be spectrally distinguished. However, the presence of WL absorption peaks observed in PLE unambiguously revealed the WL present for all studied Ga amounts. The presence of the WL was also confirmed by time-resolved measurements of the emission decay lifetime, since the WL decay lifetime was found to be an order of magnitude shorter than the free-electron acceptor decay lifetime. These measurement techniques show a clear presence of a WL signal for all investigated samples, indicating that the depositions amounts investigated, 1.4\,ML to 2.5\,ML  Ga deposition followed by 1.4\,ML In deposited at 350\,$^\circ$C did not result in the suppression of wetting layer formation during In droplet epitaxy. This is in contrast with a previous report of WL-free InAs droplet epitaxy dots \cite{Skiba17} which may indicate the effect of subtle differences in growth parameters (e.g., chamber background pressure during droplet deposition) due to the different MBE equipment used. In addition, TEM measurements demonstrated no significant change in the thickness of the WL for different deposited Ga amounts. This observation shows that the shift in the WL emission wavelength observed is mainly an effect of changing composition of InGaAs within the WL along the Ga-gradient. This work demonstrates the sensitivity of complementary PLE, TR and TEM measurements to identify the presence of a WL in the absence of a WL emission signature in photoluminescence measurements.

\ack

We thank members of the ML4Q Cluster for stimulating discussions and 
Christoph Krause and Benjamin Bennemann for technical support in growing the sample. Access to the Helmholtz Nano Facility (HNF) and the Ernst Ruska-Centre (ER-C) at Forschungszentrum Jülich is acknowledged. We acknowledge funding from Germany’s Excellence Strategy-Cluster of Excellence Matter and Light for Quantum Computing (ML4Q) EXC 2004/1-390534769.

\section*{References}
\bibliographystyle{unsrt}
\bibliography{BibFile}
	
%\listoftodos

\end{document}